\documentclass[prl,twocolumn,aps,showpacs,preprintnumbers,amsmath,amssymb]{revtex4}
\usepackage{graphicx}

\begin{document}
\preprint{}
\draft

\title{Entangling photons using a charged quantum dot in a microcavity}

\author{C.Y.~Hu$^1$}\email{chengyong.hu@bristol.ac.uk}
\author{W.J.~Munro$^2$}
\author{J.G.~Rarity$^1$}
\affiliation{$^1$Department of Electrical and Electronic Engineering, University of Bristol, University Walk, Bristol BS8 1TR, United Kingdom}
\affiliation{$^2$Hewlett-Packard Laboratories, Filton Road, Stoke Gifford, Bristol BS34 8QZ, United Kingdom}

\begin{abstract}

We present two novel schemes to generate photon polarization entanglement via single electron spins confined in charged quantum dots inside
microcavities. One scheme is via  entangled remote electron spins followed by negatively-charged exciton emissions, and another scheme is via a
single electron spin followed by the spin state measurement. Both schemes are based on giant circular birefringence and giant Faraday rotation
induced by a single electron spin in a microcavity. Our schemes are deterministic  and can generate an arbitrary amount of multi-photon
entanglement. Following similar procedures, a scheme for a photon-spin quantum interface is proposed.

\end{abstract}

\date{\today}


\maketitle

Entanglement lies at the heart of quantum mechanics and is a useful resource in quantum information sciences, especially for quantum
communications, quantum computation, quantum metrology and quantum networks \cite{nielsen00, map04}. Entanglement has been demonstrated in
various quantum systems, among which photons are well investigated as an ideal candidate to transmit quantum information and even for quantum
information processing \cite{kok07}. Starting from entangled photon pairs, multi-photon entanglement can be built, and to our knowledge the
current record is six-photon entanglement, which is limited by the brightness and photon indistinguishability of entangled photon-pair sources
\cite{lu07}. There exist three main ways to generate polarization entangled photon pairs. One way is via spontaneous parametric processes in
nonlinear crystals or fibres where non-deterministic photon pairs are created \cite{kwiat95, fulconis07}. The second way is via radiative
quantum cascades in a single atom \cite{aspect82} or biexciton in a semiconductor quantum dot (QD) \cite{benson99, stevenson06}, by which
deterministic photon pairs are generated. A third way is via single photon mixing at a non-polarizing beam splitter followed by coincidence
measurement \cite{ou88, fattal04}. However, all these methods demand quantum interference through photon or path indistinguishability, and only
one entanglement bit is obtained. An arbitrary amount of entanglement is highly desirable for quantum communications, quantum computation and
quantum networks \cite{lamata07}.

In our previous work, we proposed a novel optical non-destructive method - giant Faraday rotation - to detect a single electron spin confined in
a QD, which can be used to generate remote spin entanglement \cite{hu07}. In this Letter, we present two novel schemes to generate photon
 polarization entanglement  via single electron spins confined in charged QDs inside microcavities.  One scheme is via
entangled remote electron spins followed by negatively-charged exciton emissions, and another scheme is via a single electron spin followed by
spin state measurement. Both schemes are based on giant circular birefringence and giant Faraday rotation induced by a single electron spin. Our
schemes are deterministic  and can be used to generate an arbitrary amount of multi-photon entanglement, but it is not limited by the brightness
and photon indistinguishability of entangled photon-pair sources. Besides the spin-photon, spin-spin and photon-photon entanglement, a scheme
for a photon-spin quantum interface is proposed. These enable us to make all building blocks for quantum computation and quantum networks based
on photons and spins.

The optical properties of singly charged QDs are dominated by the optical transitions of the negatively-charged exciton ($X^-$) which consists
of two electrons bound to one hole \cite{warburton97}. Due to the Pauli's exclusion principle, $X^-$ shows spin-dependent optical transitions
\cite{hu98}: the left-handed circularly polarized photon (marked by $|L\rangle$ or L-photon) only couples the electron in the spin state
$|\uparrow\rangle$  to $X^-$ in the spin state $|\uparrow\downarrow\Uparrow\rangle$ with the two electron spins antiparallel; the right-handed
circularly polarized photon (marked by $|R\rangle$ or R-photon) only couples the electron in the spin state $|\downarrow\rangle$ to $X^-$ in the
spin state $|\uparrow\downarrow\Downarrow\rangle$. Here $|\uparrow\rangle$  and $|\uparrow\rangle$ represent electron spin states $|\pm
\frac{1}{2}\rangle$, $|\Uparrow\rangle$ and $|\Downarrow\rangle$ represent heavy-hole spin states $|\pm\frac{3}{2}\rangle$.

Now we consider a charged InGaAs/GaAs  QD inside a  micropillar microcavity with circular cross section (see Fig. 1).  The two GaAs/Al(Ga)As
distributed Bragg reflectors (DBR) and the transverse index guiding provide the three-dimensional confinement of light. The microcavity is
expected to vary or enhance the optical properties of QDs  via cavity quantum electrodynamics (cavity-QED). For simplicity, a single-sided
cavity is considered here:  the bottom DBR is highly reflective while the top DBR is partially reflective in order to couple the light into and
out of the cavity. The QD is located at the antinodes of the cavity field to achieve optimized light-matter coupling. Cavity-QED is governed by
three parameters: $\text{g}$, $\kappa$ and $\gamma$, where $\text{g}$ is the $X^-$-cavity coupling strength, $\kappa$ is twice the cavity field
decay rate, and $\gamma$ is twice the QD dipole decay rate. By solving the Heisenberg equations for the cavity field operator and QD dipole
operator in the approximation of weak excitation, we can calculate the reflection coefficient $r(\omega)\equiv |r(\omega)|e^{i\varphi(\omega)}$,
where $|r(\omega)|$ is the reflectance and $\varphi(\omega)$ is the phase shift. Details can be found in Ref. \cite{hu07}.

\begin{figure}[ht]
\centering
\includegraphics* [bb= 56 329 426 699, clip, width=5cm,height=5cm]{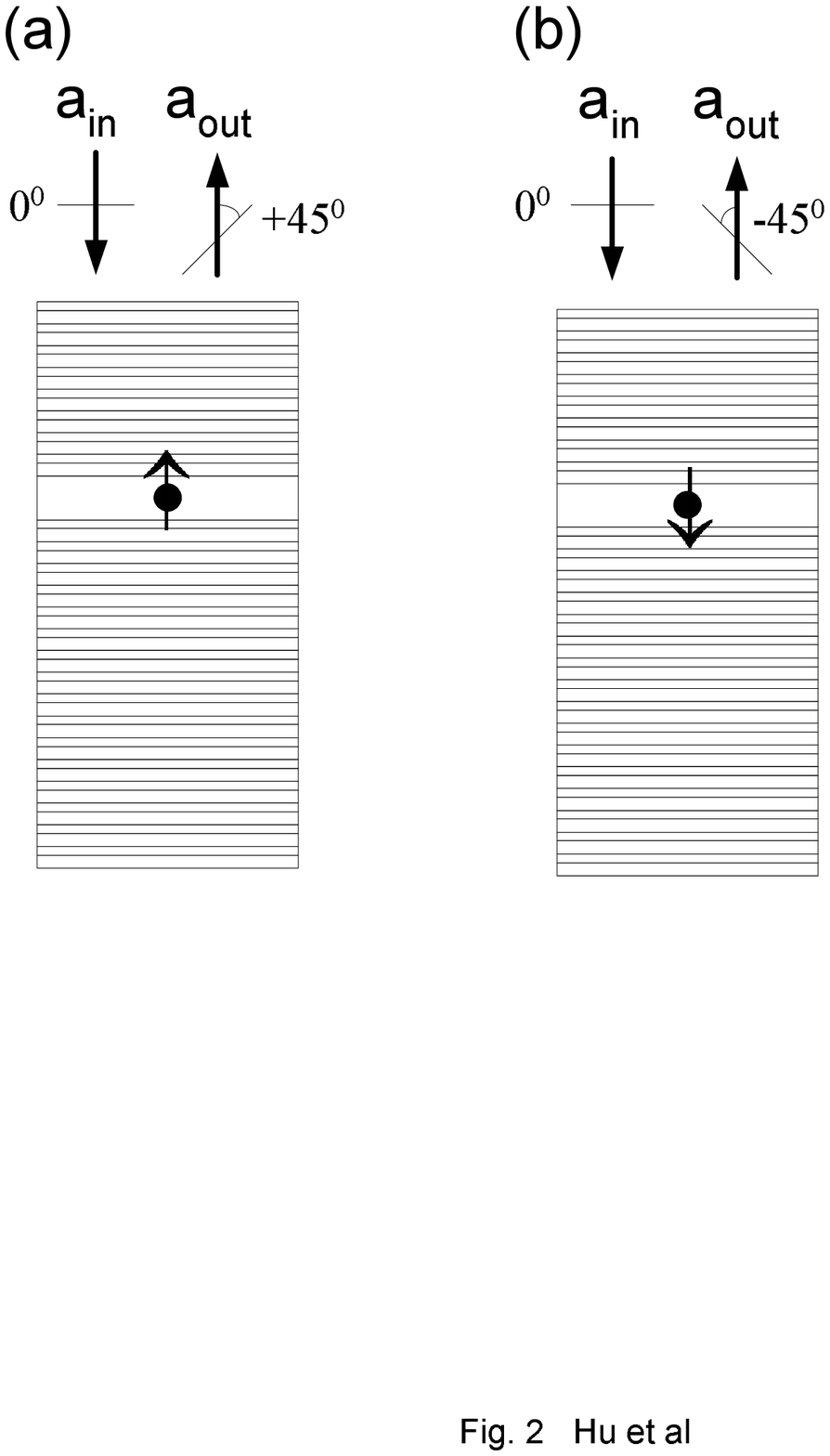}
\caption{Schematic diagram of a quantum non-demolition measurement of a single electron spin  based on giant optical Faraday rotation. The
charged QD is located at the center of  a  micropillar microcavity and strongly couples to the cavity.  In this paper, the Faraday rotation
angle is tuned to be $\pm 45^0$ (corresponding to a phase shift $\Delta \varphi=\pi/2$) by setting $\omega-\omega_c=\kappa/2$.} \label{fig1}
\end{figure}

For an empty cavity (cold cavity) without a QD inside the cavity or with a QD inside but decoupling to the cavity mode, we obtain
$|r_0(\omega)|=1$ and $\varphi_0(\omega)=\pm \pi+2\arctan 2(\omega-\omega_c)/\kappa$ where \textquotedblleft +\textquotedblright
 stands for the case of $\omega \leq \omega_c$ and \textquotedblleft -\textquotedblright  for $\omega \geq \omega_c$,
$\omega_c$ is the cavity mode frequency. $\varphi_0(\omega)$ can be tuned between $-\pi$ and $\pi$ by varying the frequency detuning
$\omega-\omega_c$. In the strong coupling region with $\text{g} \gg (\kappa, \gamma)$ (we call it hot cavity hereafter), we can get
$|r(\omega)|\simeq 1$ and $\varphi_h(\omega)\simeq 0$ for small frequency detuning $|\omega-\omega_c|\ll \text{g}$. If the single excess
electron lies in the spin state $|\uparrow\rangle$ \cite{spin}, the L-photon feels a hot cavity and gets a phase shift of $\varphi_h(\omega)$
after reflection, whereas the R-photon feels the cold cavity and gets a phase shift of $\varphi_0(\omega)$; Conversely, if the electron lies in
the spin state $|\downarrow\rangle$, the R-photon feels a hot cavity and get a phase shift of $\varphi_h(\omega)$ after reflection, whereas the
L-photon feels the cold cavity and gets a phase shift of $\varphi_0(\omega)$. We call this phenomenon giant circular birefringence (GCB), which
results in giant Faraday rotation (GFR) of linearly polarized light \cite{exp1}. Both GCB and GFR are induced by a single electron spin in a
microcavity due to cavity-QED and the optical spin selection rule of $X^-$ transitions. GFR provides a quantum non-demolition measurement of a
single electron spin (see Fig. 1), whereas  GCB can be used to make a phase gate with a phase shift operator defined as
\begin{equation}
\hat{U}(\Delta\varphi)=e^{i\Delta\varphi(|L\rangle \langle L|\otimes |\uparrow \rangle\langle \uparrow| + |R\rangle\langle R|\otimes |\downarrow
\rangle\langle \downarrow|)},
\label{gate}
\end{equation}
where $\Delta\varphi=\varphi_h(\omega)-\varphi_0(\omega)\simeq -\varphi_0(\omega)$  for small frequency detuning $|\omega-\omega_c|\ll
\text{g}$. In the following, we show that this phase gate can  be used to generate photon-photon entanglement and  build a photon-spin quantum
interface. Unless otherwise specified, we set $\Delta\varphi=\frac{\pi}{2}$ by adjusting $\omega-\omega_c=\kappa/2$ in this paper \cite{exp2}.

\begin{figure}[ht]
\centering
\includegraphics* [bb= 103 389 476 598, clip, width=6cm, height=3.5cm]{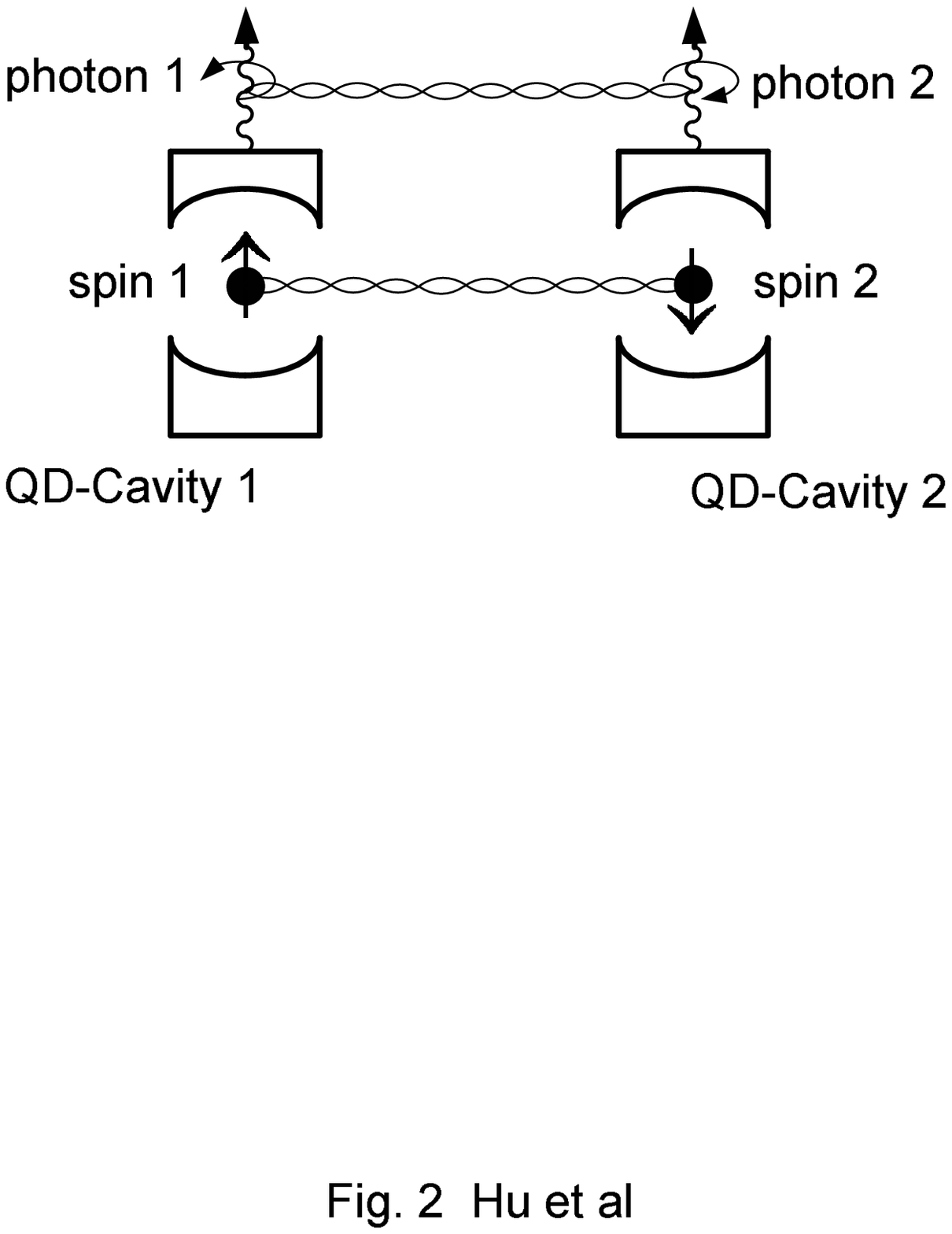}
\caption{A proposed scheme to create polarization entanglement between photons via entangled electron spins confined in  QDs.} \label{fig2}
\end{figure}

\textbf{A. Photon entanglement via entangled remote electron spins.} As discussed in our previous work \cite{hu07}, the  spin states
$|\psi^{s}_1\rangle=\alpha_1|\uparrow\rangle_1+\beta_1|\downarrow\rangle_1$ and
$|\psi^{s}_2\rangle=\alpha_2|\uparrow\rangle_2+\beta_2|\downarrow\rangle_2$ of two electrons confined in two QDs inside two micro-cavities
become entangled after interacting with  a photon in a linear polarization state, eg. $|H\rangle \equiv (|R\rangle+|L\rangle)/\sqrt{2}$. The
spin-photon interaction is described by the phase shift operator defined in Eq. (\ref{gate}) with $\Delta\varphi=\pi/2$. If detecting the photon
in the $|V\rangle \equiv (|R\rangle-|L\rangle)/\sqrt{2}$ state, the two electron spins are in an  entangled state
$|\Phi^{s}_{12}\rangle=\alpha_1\alpha_2|\uparrow\rangle_1|\uparrow\rangle_2-\beta_1\beta_2|\downarrow\rangle_1|\downarrow\rangle_2$. On
detecting the photon in the $|H\rangle$ state, the two electron spins are in  another entangled state
$|\Psi^{s}_{12}\rangle=\alpha_1\beta_2|\uparrow\rangle_1|\downarrow\rangle_2+\alpha_2\beta_1 |\downarrow\rangle_1|\uparrow\rangle_2$. Here $H$
and $V$ denote horizontal and vertical polarization of the photon.

Once entangled spins are prepared,  either optical or electrical pumping can be applied to excite $X^-$ in QDs. Due to the optical spin
selection rule of $X^-$ transitions, $X^-$ emissions yields two entangled photons in the state
\begin{equation}
|\Phi^{ph}_{12}\rangle=\alpha_1\alpha_2|L\rangle_1|L\rangle_2-\beta_1\beta_2|R\rangle_1|R\rangle_2,
\end{equation}
or
\begin{equation}
|\Psi^{ph}_{12}\rangle=\alpha_1\beta_2|L\rangle_1|R\rangle_2+\alpha_2\beta_1 |R\rangle_1|L\rangle_2,
\end{equation}
depending on the entangled spin states (see Fig. 2).

In this scheme, spin entanglement is directly transferred  to photon entanglement via $X^-$ emissions. Multi-photon entanglement can be built
either from multi-spin entanglement \cite{hu07} or from entangled photon pairs. This is a novel deterministic way to generate polarization
entangled photons from single QDs, different from the biexciton cascades \cite{benson99, stevenson06}. This scheme creates an arbitrary amount
of entanglement rather than one entanglement bit, and  relies on the long electron spin coherence so that the spin entanglement can persist
during the $X^-$ emission process. This is the case for self-assembled InGaAs/GaAs QDs: the electron spin coherence time ($\sim \mu$s) limited
by the electron spin relaxation time ($\sim$ms) \cite{kroutvar04}, is  much longer than the $X^-$ emission lifetime ($< 1$~ns) for QDs in the
cavity. Therefore, quantum coherence is still essential in the  entanglement generation.

As the QD-cavity systems work in the strong coupling regime, the photon emitting is highly efficient \cite{cui05} and the time jitter between photons
is determined by the cavity photon lifetime which is quite short.

\begin{figure}[ht]
\centering
\includegraphics* [bb= 121 441 480 584, clip, width=6cm, height=2.5cm]{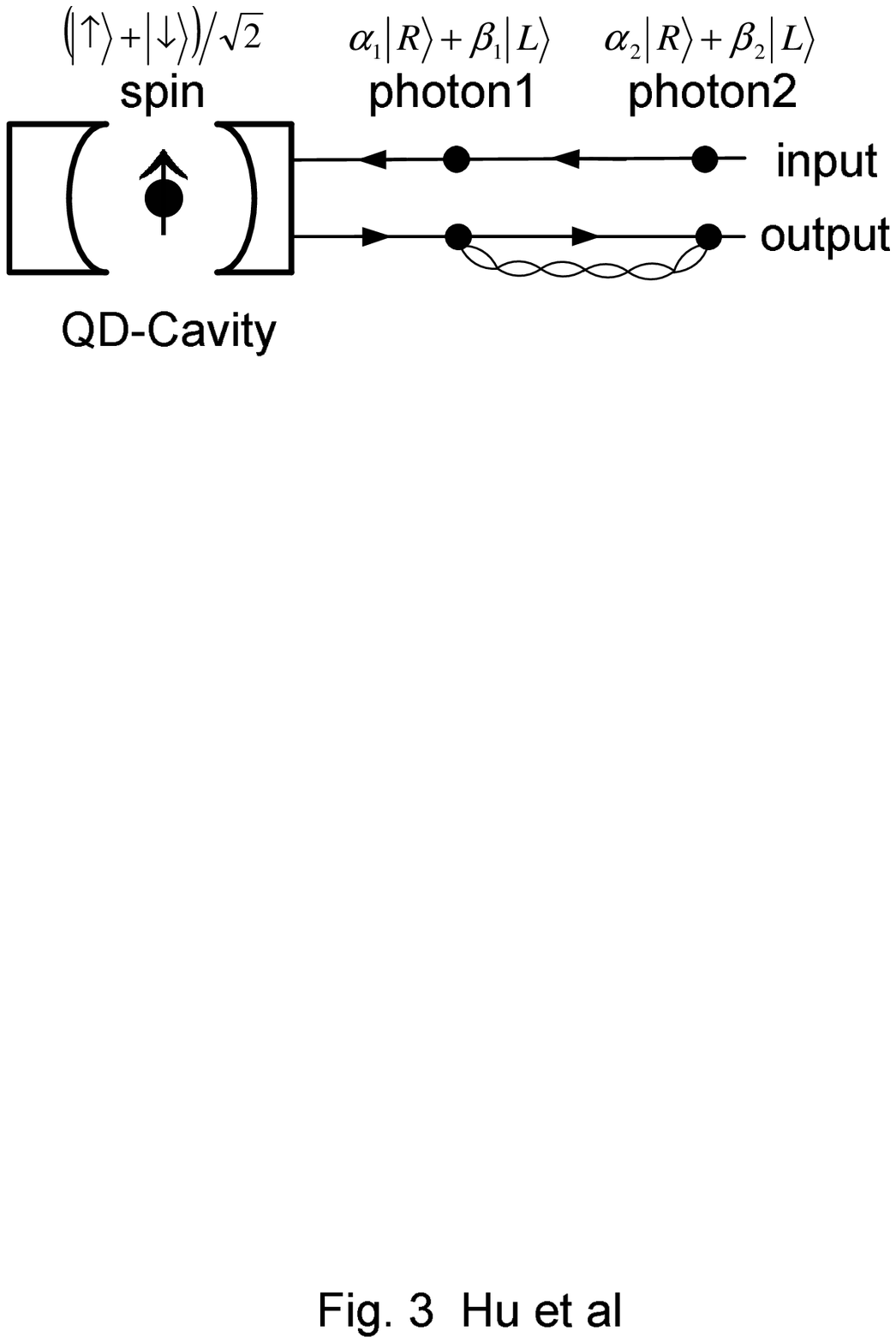}
\caption{A proposed scheme to create polarization entanglement between photons via a single electron spin confined in a QD.} \label{fig3}
\end{figure}

\textbf{B. Photon entanglement via a single electron spin.} In Fig. 3, photon 1 in the state
$|\psi^{ph}\rangle_1=\alpha_1|R\rangle_1+\beta_1|L\rangle_1$ and photon 2 in the state
$|\psi^{ph}\rangle_2=\alpha_2|R\rangle_2+\beta_2|L\rangle_2$ are input into the cavity in sequence \cite{exp3}. Both photons have the same
frequency. The electron spin in the QD is prepared in a superposition state
$|\psi^{s}\rangle=\frac{1}{\sqrt{2}}(|\uparrow\rangle+|\downarrow\rangle)$ and the phase shift operator for the QD-cavity system is  described
by Eq. (\ref{gate}) with $\Delta\varphi=\pi/2$. After reflection, the photon states become entangled with the spin state, and the corresponding
state transformation is
\begin{equation}
\begin{split}
(\alpha_1 & |R\rangle_1  +\beta_1|L\rangle_1) \otimes (\alpha_2|R\rangle_2+\beta_2|L\rangle_2) \otimes (|\uparrow\rangle+|\downarrow\rangle) \\
~ \rightarrow & ~~~ (|\uparrow\rangle-|\downarrow\rangle)\left[\alpha_1\alpha_2|R\rangle_1|R\rangle_2-
\beta_1\beta_2|L\rangle_1|L\rangle_2\right] \\
& +i(|\uparrow\rangle+|\downarrow\rangle)\left[\alpha_1\beta_2|R\rangle_1|L\rangle_2+\alpha_2\beta_1 |L\rangle_1|R\rangle_2\right].
\end{split}
\label{trans1}
\end{equation}

By applying a Hadamard gate on the electron spin (eg. using a $\pi/2$ magnetic pulse), the two spin superposition states can be rotated to the
states $|\uparrow\rangle$ and $|\downarrow\rangle$. Now the electron spin states can be measured by the GFR-based quantum non-demolition method
shown in Fig. 1. Photon 3 in the state $(|R\rangle_3+|L\rangle_3)/\sqrt{2}$ is input into the cavity (photon 3 has the same frequency as photon
1 and 2), after reflection, the total state for the three photons and one spin becomes
\begin{equation}
\begin{split}
& ~ (|R\rangle_3+i|L\rangle_3)|\uparrow\rangle\left[\alpha_1\alpha_2|R\rangle_1|R\rangle_2-
\beta_1\beta_2|L\rangle_1|L\rangle_2\right]\\
& -(|R\rangle_3-i|L\rangle_3)|\downarrow\rangle\left[\alpha_1\beta_2|R\rangle_1|L\rangle_2+\alpha_2\beta_1 |L\rangle_1|R\rangle_2\right].
\end{split}
\label{trans5}
\end{equation}

The output state of  photon 3  can be measured in orthogonal linear polarizations. If the photon 3 is detected in the $|R\rangle_3+i|L\rangle_3$
state ($45^0$ linear), so the electron spin is definitely in the state $|\uparrow\rangle$ and we project Eq. (\ref{trans5}) onto an entangled
photon pair state
\begin{equation}
|\Phi^{ph}_{12}\rangle=\alpha_1\alpha_2|R\rangle_1|R\rangle_2-\beta_1\beta_2|L\rangle_1|L\rangle_2. \label{ent1}
\end{equation}
On detecting the photon 3 in the $|R\rangle_3-i|L\rangle_3$ state ($-45^0$ linear), so the spin is  definitely in the state
$|\downarrow\rangle$ and we project Eq. (\ref{trans5}) onto another entangled photon pair state
\begin{equation}
|\Psi^{ph}_{12}\rangle=\alpha_1\beta_2|R\rangle_1|L\rangle_2+\alpha_2\beta_1 |L\rangle_1|R\rangle_2. \label{ent2}
\end{equation}

Obviously, an arbitrary amount of polarization entanglement between photons is obtained. Multi-photon entanglement can be built either by
interacting  multiple independent photons with a single electron spin, or from entangled photon pairs. Similar to  scheme A, this scheme is
deterministic and  relies on the long electron spin coherence, i.e., the time difference between photons should be shorter than the electron
spin coherence time ($\sim \mu$s) in QDs.

Scheme A and B are  both based on giant circular birefringence  and giant Faraday rotation. In the following, we show that a photon-spin
quantum interface can be built by applying the same phase gate $\hat{U}(\Delta\varphi)$ and single-spin detection method.

\begin{figure}[ht]
\centering
\includegraphics* [bb= 129 305 443 590, clip, width=5.5cm, height=5cm]{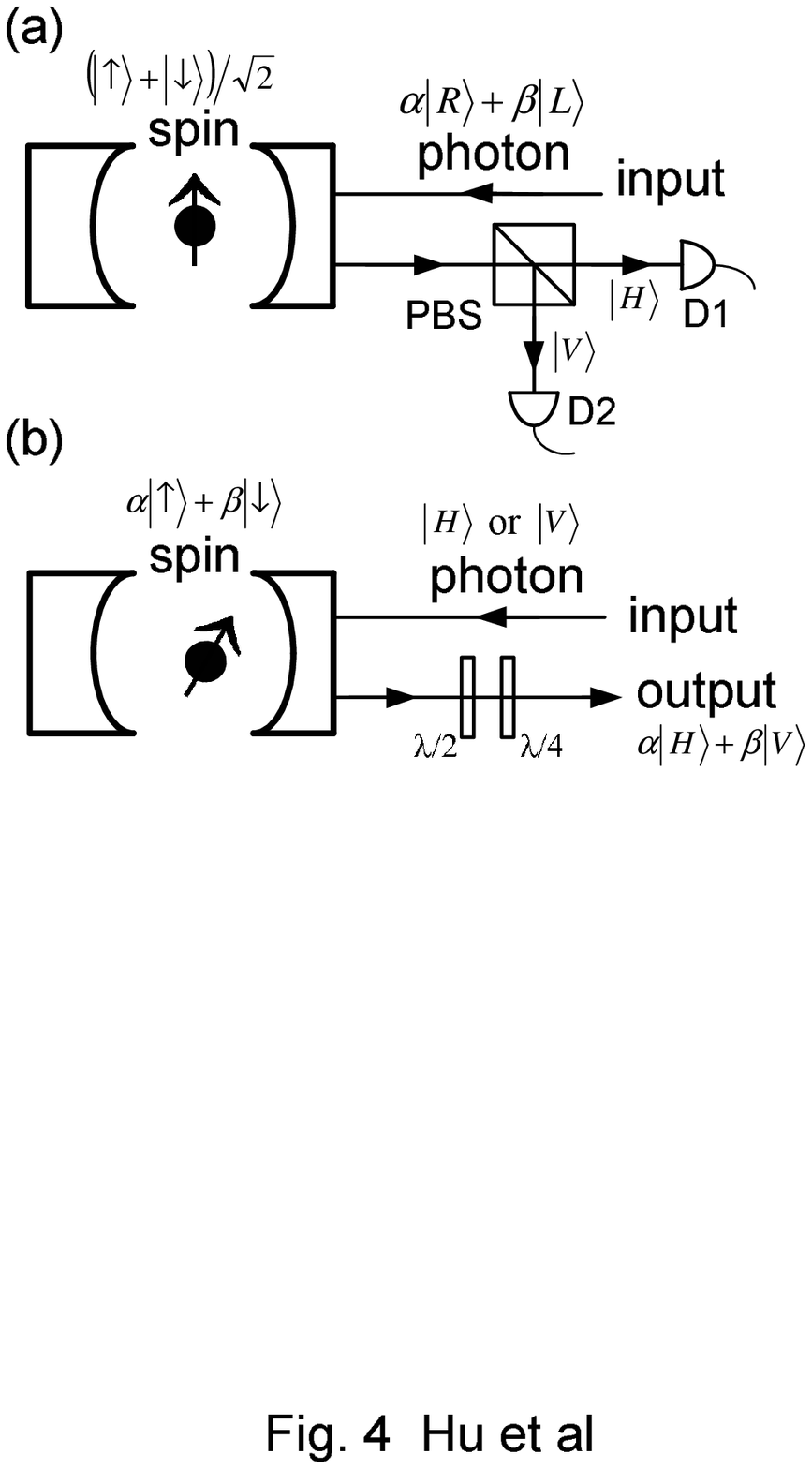}
\caption{A proposed scheme to transfer quantum state (a) from a photon to an electron spin, and (b) from an electron spin to a photon.
PBS - polarizing beam splitter; D1, D2 - photon detectors; $\lambda/2$, $\lambda/4$ - half-, quarter-wave plates. } \label{fig4}
\end{figure}

\textbf{C. Quantum state transfer from photon to spin.} In Fig. 4(a), a photon in a unknown state
$|\psi^{ph}\rangle=\alpha|R\rangle+\beta|L\rangle$ is input to a QD-cavity system with the electron spin prepared in the state
$|\psi^{s}\rangle=\frac{1}{\sqrt{2}}(|\uparrow\rangle + |\downarrow\rangle)$, and a phase shift operator $\hat{U}(\Delta\varphi=\pi/2)$.
After reflection, the photon state and spin state become entangled, i.e,
\begin{equation}
\begin{split}
&(\alpha|R\rangle+\beta|L\rangle) \otimes (|\uparrow\rangle + |\downarrow\rangle)  \rightarrow \\
&  \alpha |R\rangle |\uparrow\rangle + i\alpha|R\rangle |\downarrow\rangle +
i\beta|L\rangle |\uparrow\rangle+\beta |L\rangle|\downarrow\rangle.
\end{split}
\label{trans6}
\end{equation}
By applying a Hadamard gate on the photon state (eg. using a polarizing beam splitter), we obtain a spin state
$|\Phi^s_1\rangle=\alpha(|\uparrow\rangle+i|\downarrow\rangle)+i\beta(|\uparrow\rangle-i|\downarrow\rangle)$ if detecting a photon in the
$|H\rangle$ state, or a spin state $|\Psi^s_1\rangle=\alpha(|\uparrow\rangle+i|\downarrow\rangle)-i\beta(|\uparrow\rangle-i|\downarrow\rangle)$
if detecting a photon in the $|V\rangle$ state. Next, by applying a Hadamard gate on the electron spin (eg. using a $\pi/2$ magnetic pulse), the
electron spin state is converted to $|\Phi^s_1\rangle=\alpha|\uparrow\rangle+i \beta|\downarrow\rangle$ and
$|\Psi^s_1\rangle=\alpha|\uparrow\rangle-i \beta|\downarrow\rangle$. By applying a unitary operation on the electron spin states, we get
 $|\psi^s\rangle=\alpha|\uparrow\rangle+\beta|\downarrow\rangle$.
Therefore, a unknown photon state is transferred to the electron spin state.

\textbf{D. Quantum state transfer from spin to photon.} In Fig. 4(b), a photon in the polarization state
$|\psi^{ph}\rangle=(|R\rangle+|L\rangle)/\sqrt{2}$ is input to a QD-cavity system with the electron spin in a unknown state
$|\psi^{s}\rangle=\alpha |\uparrow\rangle+\beta |\downarrow\rangle$, and a phase shift operator $\hat{U}(\Delta\varphi=\pi/2)$.
After reflection, the photon state and spin state become entangled, i.e,
\begin{equation}
\begin{split}
&(|R\rangle+|L\rangle) \otimes (\alpha |\uparrow\rangle+ \beta |\downarrow\rangle) \rightarrow \\
& \alpha(|R\rangle+i|L\rangle)|\uparrow\rangle+\beta (i|R\rangle+|L\rangle)|\downarrow\rangle.
\end{split}
\label{trans7}
\end{equation}
After  applying a Hadamard gate on the electron spin (eg. using a $\pi/2$ magnetic pulse), a third photon in the state
$(|R\rangle+|L\rangle)/\sqrt{2}$ is input to measure the electron spin state by using the GFR-based quantum non-demolition
measurement (see Fig. 1). If detecting the electron spin in the  $|\uparrow\rangle$ state, the photon is then projected in the state
$|\Phi^{ph}_1\rangle=\alpha|+45^0\rangle+i\beta|-45^0\rangle$; If detecting the electron spin in the  $|\downarrow\rangle$ state,
the photon is then projected in the state $|\Psi^{ph}_1\rangle=\alpha|+45^0\rangle-i\beta|-45^0\rangle$. Here $|\pm45^0\rangle\equiv(|R\rangle\pm
i|L\rangle)/\sqrt{2}$. By applying a unitary operation (eg. using a $\lambda/2$ and a $\lambda/4$ wave plate) on the photon state,
we get $|\psi^{ph}\rangle=\alpha|H\rangle+\beta|V\rangle$ \cite{exp3}. So a unknown spin state is transferred to the photon state.

Different from the original teleportation protocol which involves three qubits \cite{bennett93}, our state-transfer scheme needs
only two qubits thanks to  an  arbitrary amount of spin-photon entanglement achieved. The state-transfer fidelity is affected by
the imperfect reflectance of the hot cavity, i.e., $|r(\omega)|\rightarrow 1$ when $|\omega-\omega_c| \ll \text{g}$, and the
asymmetric shape of QDs in practice.

In  conclusion, we have presented two novel schemes to generate an arbitrary amount of photon polarization entanglement via single electron
spins confined in charged QDs inside microcavities based on the giant circular birefringence and giant Faraday rotation induced by a single
electron spin. Both schemes are deterministic and rely on the electron spin coherence, indicating that quantum coherence remains essential to
the entanglement generation. Our schemes can be used to generate multi-photon entanglement, but it is not limited by  the brightness and photon
indistinguishability of entangled photon-pair sources. A scheme of a quantum interface between photon and spin is also proposed. With the
spin-photon, spin-spin and photon-photon entanglement as well as  the photon-spin quantum interface, we can make all building blocks, such as
quantum memories, quantum repeaters and various quantum logic gates for quantum communications, quantum computation and quantum networks. We
believe this work opens a new avenue in quantum information sciences.

J.G.R is supported by a Wolfson merit award.  This work is partly funded by UK EPSRC IRC in Quantum Information Processing
and QAP (EU IST 015848).


\begin{references}
\bibitem{spin}The spin is quantized along the propagation direction of the input (or output) photons.
\bibitem{exp2}Single photons generated from self-assembled InGaAs QDs have the spectral linewidth much smaller than
the cavity broadening, i.e., $\gamma \ll \kappa$. So the frequency detuning $\omega-\omega_c=\kappa/2$ can be precisely achieved.
\bibitem{exp3}By using a $\lambda/4$ wave plate, the photon state $|\psi^{ph}\rangle=\alpha|R\rangle+\beta|L\rangle$ can be converted
to $|\psi^{ph}\rangle=\alpha|H\rangle+\beta|V\rangle$ and vice versa.
\end{references}
\end{document}